# The Mass Function of Cosmic Structures with Non-spherical Collapse

Pierluigi Monaco

Scuola Internazionale Superiore di Studi Avanzati (SISSA), via Beirut 4, 34014 – Trieste, Italy
Dipartimento di Astronomia, Università degli studi di Trieste, via Tiepolo 11, 34131 – Trieste, Italy
email: monaco@tsmi19.sissa.it

**Abstract**

Non-spherical dynamical approximations and models for the gravitational collapse are used to extend the well-known Press & Schechter (PS) approach, in order to determine analytical expressions for the mass function of cosmic structures. The problem is rigorously set up by considering the intrinsic Lagrangian nature of the mass function. The Lagrangian equations of motion of a cold and irrotational fluid in single-stream regime show that the shear, which is non-locally determined by all the matter field, is the quantity which characterizes non-spherical perturbations. The Zel'dovich approximation, being a self-consistent first-order Lagrangian and local one, is used as a suitable guide to develop realistic estimates of the collapse time of a mass clump, starting from the local initial values of density and shear. Both Zel'dovich-based *ansätz* and models and the homogeneous ellipsoidal model predict that more large-mass objects are expected to form than the usual PS relation. In particular, the homogeneous ellipsoid model is consistent at large masses with a Press & Schechter mass function with a lower value of the $\delta_c$ parameter, in the range $1.4 \div 1.6$. This gives a dynamical explanation of why lower $\delta_c$ values have been found to fit the results of several N-body simulations. When more small-scale structure is present, highly non-linear dynamical effects can effectively slow down the collapse rate of a perturbation, increasing the effective value of $\delta_c$. This may have interesting consequences on the abundance of large-mass high-redshift objects.

*Subject headings*: cosmology: theory — galaxies: clustering — large-scale structure of the universe

# 1  Introduction

A fundamental feature of our observable Universe is the presence of collapsed clumps of matter, such as galaxies, groups and clusters of galaxies, whose mass distribution (mass function, hereafter MF), which has recently been observationally estimated (see, e.g, Ashman, Salucci & Persic 1993, for galaxies; Pisani et al. 1992, for groups; Henry & Arnaud 1991, for X-ray clusters; Bahcall & Cen 1993, and Biviano et al. 1993, for optical clusters), is a relevant quantity for comparison with the predictions of cosmological models of structure formation. Much effort has been focused on the problem of finding the theoretical MF of a given model; this is not an easy task, as the equations of motion of matter in the collapsing phases are non-local and highly non-linear.

A first attempt to estimate the MF of collapsed structures has been made by Press & Schechter (1974, hereafter PS) by means of very simplified assumptions: given a smoothed initial gaussian density field in a flat Universe, every mass element is supposed to follow spherical collapse (e.g., Peebles 1980), according to which infinite density is reached when the density contrast $\delta$, linearly extrapolated to the collapse time, is 1.69; in this context, the dynamical fate of a perturbation is dictated simply by its initial density contrast. It is then supposed that the collapsed element forms a structure with mass equal or greater than the mass associated to the smoothing volume adopted, so that merging of collapsing mass elements into larger ones is allowed. This approach has three main weak points: (i) the dynamical model of spherical collapse is unrealistic in general cases. (ii) The PS mass function fails to predict the collapse of clumps which are not overdense enough to collapse by themselves, but nonetheless are going to be accreted onto larger structures (*cloud-in-cloud* problem); then, to reach the correct normalization (all matter collapsed at some scale), one has to multiply the MF by a "fudge factor" 2. (iii) Every mass element evolves independently from the others, so that coagulation and fragmentation[1] of collapsed clumps are neglected.

Weak point (i) has not generally been faced: most of the results on the MF and on the abundance of structures, such as rich clusters of galaxies or quasars, rely on spherical collapse, which is asymptotically accurate only for structures coming from the rare highest peaks of the primordial density field (e.g. Bernardeau 1994). Sometimes the homogeneous ellipsoid model has been used to describe supercluster dynamics (e.g. White & Silk 1979, Hoffman 1986). More recently, Bond & Myers (1993a,b) have used the ellipsoidal model to describe the collapse of peaks, while Bartlemann, Ehlers & Shneider (1993) have used both ellipsoidal collapse and Zel'dovich approximation to estimate the collapse time of a mass element.

On the other hand, points (ii) and (iii) have been faced by many authors. The normalization "fudge factor" 2 has been justified by Peacock & Heavens (1990) and Bond et al. (1991) by means of the excursion set theory; briefly, collapse prediction and mass assignment have to be done on the basis of the largest collapsing structure which includes the given initial point. However, the resulting MF is the same as the PS one only in the case of sharp $k$-space smoothing. Furthermore, coagulation and fragmentation of clumps have been described by means of Smoluchowsky equation (Silk & White 1978; Lucchin 1988; Cavaliere, Colafrancesco & Menci 1991; Cavaliere & Menci 1994), with the result that at small scales, or in some special kind of environment (e.g. cD galaxies in rich clusters), the MF can lose memory of the cosmologically relevant initial conditions. Finally, Cavaliere & Menci (1994), by means of Cayley trees formalism, have succeeded in unifying the excursion set and Smoluchowsky equation approaches.

Full N-body calculations have surprisingly shown that the PS mass function gives reasonable fits, despite of its weak points (Efstathiou et al. 1988; Efstathiou & Rees 1988; Carlberg & Couchman 1989; Bond et al. 1991; Bond & Myers 1993b; Lacey & Cole 1994; Klypin et al. 1994); however, the comparisons between PS and N-body MFs have been made only in limited ranges of masses, and some tuning of the MF parameters has been needed to achieve good fits. At variance with the previous authors, Jain & Bertschinger (1993) have claimed that, for a standard cold dark matter spectrum, the PS mass function significantly underestimates the number of high-redshift objects, while it gives a reasonable fit at later times, consistently with their calculations on the non-linear growth of perturbations. They have ascribed the difference between their result and the previous ones to the improved resolution of their N-body $P^3M$ code.

---

[1] Here two mass elements *merge* if they are part of a larger-scale collapsing clump, *coagulate* if their interaction is rather a two-body collision between already collapsed clumps; coagulation and fragmentation are responsible for that part of the merging histories which can not be predicted from local collapse and mass assignment criteria.

The PS approach, originally limited to an Einstein-de Sitter Universe, has been extended to the more general cases of open Universes with non-vanishing cosmological constant (Lilje 1992, and references therein), and of non-Gaussian statistics (Lucchin & Matarrese 1988). Other approaches have been used to determine an analytical expression for the MF. Within the peak paradigm, according to which structures form in the peaks of the initial density field, the density of peaks (Bardeen et al. 1986) can be used to construct a MF; however, it is not clear which mass has to be assigned to a peak and, moreover, the peak paradigm has been put under discussion by recent N-body works (Katz, Quinn & Gelb 1993; van de Weygaert & Babul 1994; but see, for a different view, Bond & Myers 1993a,b). Finally, Cavaliere, Colafrancesco & Scaramella (1991) have developed a dynamical approach to the mass function, explicitly introducing time-scales for creation and destruction of clumps.

To deepen the knowledge on the MF it is necessary to consider more realistic models for the collapse of a mass element than the spherical one. In this paper I consider dynamical models and approximations for the non-linear evolution of perturbations which are able to predict the fate of a mass element from local initial conditions; in this case it is possible to extende the original PS approach, which seems to offer the best compromise between simplicity and success. In §2 the intrinsic Lagrangian nature of the MF is stressed, so that the Zel'dovich approximation, which is local, provides a natural and useful guide to estimate the collapse time of a mass element. In §3 a correction to the PS mass function is defined and some *ansätz* for the collapse time, based on the Zel'dovich approximation, are introduced and their consequences on the MF are examined; then, two more realistic models are analyzed: one based on the Zel'dovich approximation, and the other on the homogeneous ellipsoid model. In §4 the physical differences between the usual PS mass function and the new ones here proposed are outlined and discussed, and in §5 the results are summarized and final conclusions are drawn.

## 2 The Mass Function As A Lagrangian Quantity

To assess in a rigorous way the problem of finding an approximation for the collapse time, it is useful to start with the equations for the evolution of a cosmological fluid. What is relevant to our purpose is the density of a given mass element — the Lagrangian density — and, in particular, the instant at which it diverges, which is the collapse time. Then, the MF can be regarded as an intrinsically Lagrangian quantity. It is useful to restrict ourselves to a matter-dominated flat Universe with vanishing cosmological constant, filled with pressureless and non-vortical matter; moreover, the initial field is assumed to be linear, Gaussian, and growing-mode dominated, and the dynamical description is limited to the single-stream regime (no orbit crossing having occurred yet). These assumptions, which are shared by the usual PS approach, can be relaxed to more general cosmologies, retaining the validity of the qualitative features of the present results.

A convenient form for the Lagrangian Newtonian evolution equations of a cosmological fluid can be obtained (Gurbatov, Saichev & Shandarin 1989; Matarrese et al. 1992) by using the scale factor $a$ (which is also the growth factor of the linear growing mode) as a time variable ($a$ is normalized to be unity at the present time). In this way, being $\mathbf{x}$ the Eulerian coordinate, the velocity of the fluid element is $\mathbf{u} \equiv d\mathbf{x}/da = \dot{\mathbf{x}} = \mathbf{v}/a(da/dt)$ ($\mathbf{v}$ is the usual peculiar velocity), $d/da \equiv \partial/\partial a + \mathbf{u} \cdot \nabla_{\mathbf{x}}$ is the Lagrangian time derivative, the density contrast $\delta(\mathbf{q})$ is defined as usual as $[\varrho(\mathbf{q}) - \bar{\varrho}]/\bar{\varrho}$, and the potential $\varphi$ is defined as $\varphi \equiv 3t_0^2/2a_0^3 \, \phi$, where $\phi$ is the usual gravitational potential. With these definitions, the Euler, continuity and Poisson equations become:

$$\dot{\mathbf{u}} + \frac{3}{2a}\mathbf{u} = -\frac{3}{2a}\nabla_{\mathbf{x}}\varphi \qquad (1)$$

$$\dot{\delta} + (1+\delta)\nabla_{\mathbf{x}} \cdot \mathbf{u} = 0 \qquad (2)$$

$$\nabla_{\mathbf{x}}^2 \varphi = \frac{\delta}{a} \qquad (3)$$

These equations are valid as long as the fluid is in single-stream regime.

The Eulerian gradients of the components of the peculiar velocity $\mathbf{u}$ can be decomposed, without loss of generality, into the expansion scalar $\vartheta$, the shear tensor $\Sigma$ and the vorticity tensor $\omega$:

$$\nabla_i u_j = \frac{1}{3}\vartheta \delta_{ij} + \Sigma_{ij} + \omega_{ij} \tag{4}$$

(see, e.g., Ellis, 1971). $\vartheta$ is simply the divergence of **u**, while $\Sigma_{ij}$ and $\omega_{ij}$ are the traceless symmetric and antisymmetric parts of $\nabla_i u_j$. $\omega_{ij}$ is of course null for an irrotational fluid. Whit these kinematical quantities it is possible to find the evolution equation for the density contrast:

$$\ddot{\delta} + \frac{3}{2a}\dot{\delta} = \frac{4}{3}\frac{\dot{\delta}^2}{(1+\delta)} + (1+\delta)\left(\Sigma^2 - \omega^2 + \frac{3}{2a}\delta\right) \tag{5}$$

where $\Sigma^2 \equiv \Sigma_{ij}\Sigma_{ij}$ and $\omega^2 \equiv \omega_{ij}\omega_{ij}$. In the case $\Sigma = \omega = 0$, Eq. (5) describes the evolution of a spherical perturbation, and its solutions are the well-known ones of the spherical collapse model used in the PS mass function. It is clear that, while the $\omega = 0$ hypothesis is quite reasonable in a cosmological context, at least before orbit crossing (e.g. Peebles 1980), the hypothesis of vanishing shear is unrealistic for a general perturbation. It is to be noted that the shear-dependent term in Eq (5), $\Sigma^2$, is positive-definite, so that, as long as the fluid is irrotational, the growth rate of the density contrast of a mass element is always enhanced by the shear. As a consequence, a general mass element always collapses faster than an equally overdense spherical one. This fact, already noted by Hoffman (1986), has been put in form of theorem by Bertschinger & Jain (1994). Moreover, the shear term $\Sigma^2$ is the one which introduces the non-locality of Poisson equation into Eq. (5); indeed, the evolution equation for $\Sigma_{ij}$ is :

$$\dot{\Sigma}_{ij} + \frac{2}{3}\vartheta\Sigma_{ij} + \Sigma_{il}\Sigma_{lj} + \frac{3}{2a}\Sigma_{ij} - \frac{1}{3}\Sigma^2 \delta_{ij} = -\frac{3}{2a}\left(\nabla_i\nabla_j\varphi - \frac{1}{3}\nabla^2\varphi\delta_{ij}\right), \tag{6}$$

where the source term between parentheses on the rhs is the tidal field tensor, which is non-local (e.g., Kofman & Pogosyan 1994). Non-local estimates of the collapse time make the PS approach inappropriate and introduce strong computational difficulties; then, in order to study the effect of non-spherical collapse with a PS-like approach, only local estimates of the collapse time will be considered in the present work.

A number of recent papers have focused on the fact that the GR Lagrangian evolution equations of a pressureless irrotational fluid are local if the so-called magnetic part $H_{\mu\nu}$ of the Weil tensor is neglected (Barnes & Rowlingson 1989; Matarrese, Pantano & Saez 1993; Bertschinger & Jain 1994). As the Newtonian limit of that tensor is null, it can seem natural to extrapolate this approximation to that limit; however, non-null post-Newtonian terms from $H_{\mu\nu}$ enter in the Newtonian equations of motion introducing non-locality (Matarrese, Pantano & Saez 1994; Kofman & Pogosyan 1994), so that this $H_{\mu\nu} = 0$ approximation is not valid in general but only for some special symmetries (Bertschinger & Jain 1994).

Anyway, it is possible to find well-founded approximations which give a local description of the evolution of a mass clump. In particular, the most famous approximation which can be introduced in the Lagrangian framework, the Zel'dovich one (Zel'dovich 1970, hereafter ZEL; see also Shandarin & Zel'dovich 1989) is local. Let **q** and **x** be the Lagrangian and Eulerian coordinates of a fluid element, with $\mathbf{x}(\mathbf{q}, a_0) = \mathbf{q}$; the trajectory of the element is:

$$\mathbf{x}(\mathbf{q}, a) = \mathbf{q} + \Psi(\mathbf{q}, a), \tag{7}$$

where $\Psi$ is called displacement field. ZEL consists in expressing the latter as the product of a universal time factor $b(a)$ and a constant field $\mathbf{u}(\mathbf{q})$:

$$\mathbf{x}(\mathbf{q}, a) = \mathbf{q} + b(a)\mathbf{u}(\mathbf{q}); \tag{8}$$

here $b$ is the growth factor of the linear growing mode, $b(a) = a - a_0$ (in a flat Universe), and the field $\mathbf{u}(\mathbf{q})$ is the initial peculiar velocity of the fluid element. We will refer to Eq. (8) as ZEL mapping. According to it, the fluid elements maintain their initial peculiar velocity ($\dot{\mathbf{u}} = 0$), performing straight paths in the Eulerian comoving space. In this way, all physical quantities can be derived from the local values of the initial deformation tensor, i.e. the Lagrangian derivatives of the peculiar velocity field, $\partial u_i/\partial q_j$.

At a certain time fluid elements coming from different Lagrangian positions get to the same Eulerian point (orbit crossing or shell crossing); there caustics (i.e. regions of infinite density) form, and ZEL mapping becomes multi-valued, i.e. multi-stream regions form. In this case ZEL predicts the particles just to continue their straight walks, while in the real collapse the particles will form a structure. I have chosen to identify the collapse time with the first shell-crossing, as it is the instant at which the density of the mass element diverges; moreover, shock waves in the baryonic component are predicted to form at that moment (Zel'dovich 1970). This point is discussed in a deeper way at the end of §4.

ZEL can be also obtained as the first-order term of the expansion of the displacement field $\Psi$ in terms proportional to $b(a)^n$ (e.g., Buchert 1994 and references therein): it is a true self-consistent first-order Lagrangian approximation. The fact that the series expansion of the displacement field and not of the density contrast is made implies that ZEL, together with higher-order Lagrangian approximations, is not *a priori* limited to $\delta \ll 1$; as a matter of fact, ZEL is an exact solution in 1D, is accurate up to $\delta \sim 1$, and gives answers in the right direction also in the highly non-linear regime up to shell crossing (e.g., it does predict caustics to form, while the usual Eulerian linear approximation does not). Higher-order terms of the displacement field can be calculated, and their convergence toward the exact solution is assured up to orbit crossing. However, higher-order Lagrangian approximations are non-local, so they will not be considered here.

In the realistic case of matter fields which have power on all scales, so that multi-stream regions can always be found at small scales, to apply ZEL it is first necessary to smooth the initial density field, or, in other words, to truncate the power spectrum at some wavenumber (Coles, Melott and Shandarin 1993). This truncated version of ZEL has been found, by means of cross-correlation tests, to predict in a satisfactory way the relevant large-scale features of N-body simulations for spectral indexes $n$ not greater than 1 (Coles et al. 1993; Melott, Pellman & Shandarin 1993; see also Melott 1994 and references therein). Moreover, the less small-scale structure is present, the best truncated ZEL works; gaussian smoothing optimizes its performances. Furthermore, peaks in ZEL-evolved fields have been shown to describe fairly well cluster correlations (Mann, Heavens & Peacock 1993; Borgani, Coles & Moscardini 1993). Other authors (Grinstein & Wise 1993; Bernardeau et al. 1993, Munshi & Starobinsky 1993, Munshi, Sahni & Starobinsky 1994), by means of perturbative calculations, have shown that ZEL underestimates the higher moments of the evolved density field (i.e. the skewness, kurtosis etc.), but nonetheless it works better than other approximations schemes.

Other approximations for the non-linear evolution of density perturbations have been proposed, but they are generally non-local. For the adhesion model (see, e.g., Shandarin & Zel'dovich 1989), in which particles whose trajectories try to intersect simply stick together, Vergassola et al. (1994) have analytically calculated the MF with the aid of several mathematical tools borrowed from diffusion theory, finding a large-mass behaviour similar to the PS one, but a different small-mass slope. Other approximation schemes, such as frozen flow (Matarrese et al. 1992) and frozen potential (Brainerd, Sherrer & Villumsen 1993; Bagla & Padmanabhan 1994) approximations, are intrinsically Eulerian, so that particle trajectories have to be found to get any dynamical prediction for a given fluid element. In the following, the self-consistent ZEL approximation will be considered as a guide to realistic estimates of the collapse time.

## 3 Local Approximations for the Collapse Time

### 3.1 A PS-like ZEL-based Approach to the MF

Given a local estimate of the collapse time and the statistical probability distribution function (PDF) of the initial conditions, the PS approach gives a recipe to construct an analytical expression for the MF. As a first step, ZEL is used to approximate the dynamics of a mass element. It has been shown before that the Lagrangian peculiar velocity of a fluid element does not change in time, so that:

$$\mathbf{u}(\mathbf{q}) = -\nabla_{\mathbf{q}} \varphi_0, \tag{9}$$

where $\varphi_0$ is the initial peculiar gravitational potential. If we call $\lambda_1$, $\lambda_2$ and $\lambda_3$ the three eigenvalues of the deformation tensor $\partial u_i/\partial q_j$ (or, equivalently, of the tensor $-\partial^2 \varphi_0/\partial q_i \partial q_j$), assuming $\lambda_1 \geq \lambda_2 \geq \lambda_3$, the Jacobian matrix $J_{ij}$ of the ZEL mapping, Eq. (8), can be written as:

$$J_{ij} \equiv \frac{\partial x_i}{\partial q_j} = \delta_{ij} + b(a)\frac{\partial u_i}{\partial q_j} = \text{diag}\,(1 + b\lambda_1, 1 + b\lambda_2, 1 + b\lambda_3)\,. \tag{10}$$

Equation (9) (9) and Poisson Eq. (3) imply $\lambda_1 + \lambda_2 + \lambda_3 = -\delta_0/a_0$; then the three eigenvalues $\lambda_i$ give a full description of the initial conditions for the local evolution of a density perturbation. However, it is worth noting that these $\lambda_i$ eigenvalues are not local, in the sense that they are non-locally determined by the initial tidal field. This means that, at variance with the spherical top-hat model, which assumes that the Universe is homogeneous outside the perturbation, our mass elements are extracted from a perturbed density field and their dynamics is influenced, at least through the initial conditions, by the rest of the Universe.

It is very easy at this point to explicitly calculate the expansion and the shear of a mass element:

$$\begin{aligned}\vartheta &= (\mu_1 + \mu_2 + \mu_3) \\ \Sigma_{ij} &= \text{diag}\,(\mu_1 - 1/3\vartheta, \mu_2 - 1/3\vartheta, \mu_3 - 1/3\vartheta)\end{aligned} \tag{11}$$

where $\mu_i \equiv \lambda_i/(1 + a\lambda_i)$, $i$=1,2,3 (note that it has been assumed $b(a) = a - a_0 \simeq a$).

There are two different ways to define a density with ZEL (Shandarin, Doroshkevich & Zel'dovich 1983). First, the continuity Eq. (2) can be used to construct a density which is consistent with mass conservation; recalling that the Jacobian determinant represents the volume of a mass element, it is easy to obtain:

$$(1 + \delta_c) = (1 + \delta_0)|\det J^{-1}| = \frac{1 + \delta_0}{(1 + a\lambda_1)(1 + a\lambda_2)(1 + a\lambda_3)} \tag{12}$$

This continuity density $\delta_c$ is not consistent with momentum conservation; we can define a dynamical density $\delta_d$ by means of the Poisson Eq. (3), and by means of Eq. (9): $\delta_d = -a\vartheta$. In this way:

$$\delta_d = -a(\mu_1 + \mu_2 + \mu_3). \tag{13}$$

$\delta_d$ is consistent with momentum conservation but not with mass conservation. All these quantities, i.e. the expansion, the shear and the two densities defined above, diverge when $1 + a\lambda_3 = 0$, where $\lambda_3$ is the smallest of the $\lambda_i$. Then the collapse time $a_c$ can be easily written as:

$$a_c = -\frac{1}{\lambda_3} \tag{14}$$

Given this local estimate of the collapse time, to calculate the MF we need to know the PDF of the initial conditions, i.e. of the three eigenvalues of the deformation tensor $\lambda_i$. This has been calculated for a Gaussian field by Doroshkevich (1970):

$$\mathcal{P}(\lambda_1, \lambda_2, \lambda_3) = \frac{675\sqrt{5}}{8\pi\sigma^6}\exp\left(-\frac{3}{\sigma^2}s_1^2 + \frac{15}{2\sigma^2}s_2\right)(\lambda_1 - \lambda_2)(\lambda_2 - \lambda_3)(\lambda_1 - \lambda_3); \tag{15}$$

where $\sigma$ is the usual mass variance linearly extrapolated to the present time, $s_1 = \lambda_1 + \lambda_2 + \lambda_3$ and $s_2 = \lambda_1\lambda_2 + \lambda_2\lambda_3 + \lambda_1\lambda_3$. It is convenient to perform the following change of variables:

$$\begin{cases}\delta &= -\lambda_1 - \lambda_2 - \lambda_3 \\ x &= \lambda_1 - \lambda_2 \\ y &= \lambda_2 - \lambda_3\end{cases} \tag{16}$$

where $\delta$ is the density contrast linearly extrapolated to the present, and varies from $-\infty$ to $\infty$. The variables $x$ and $y$ have been defined in order that their ranges of variability are decoupled: the two inequalities $\lambda_2 \leq \lambda_3$ and $\lambda_1 \leq \lambda_2$ implicate the conditions $x \geq 0$ and $y \geq 0$. They are simply related to the shear tensor: it is easy to show that the initial eigenvalues of $\Sigma_{ij}$ are $2x/3 + y/3$, $-x/3 + y/3$ and $-x/3 - 2y/3$. Then, if $x = 0$ and $y = 0$ the shear is null and the collapse is spherical.

With the transformation (16) the PDF becomes:

$$\mathcal{P}(\delta,x,y) = \frac{1}{\sqrt{2\pi}\sigma}\exp\left(-\frac{\delta^2}{2\sigma^2}\right) \times \frac{225}{4}\sqrt{\frac{5}{2\pi}}\frac{1}{\sigma^5}\exp\left(-\frac{5}{2\sigma^2}(x^2+xy+y^2)\right)xy(x+y) = \mathcal{P}(\delta)\times\mathcal{P}(x,y) \quad (17)$$

Equation (17) is just the product of two normalized distribution: the one for $\delta$, $\mathcal{P}(\delta)$, which is the usual Gaussian distribution, and the one for $x$ and $y$, $\mathcal{P}(x,y)$.

With these new variables the ZEL collapse time becomes:

$$a_c = \frac{3}{\delta + x + 2y}. \quad (18)$$

It is useful to define a function $\delta_c(x,y)$ as the density contrast needed by a perturbation to collapse at $a_c = 1$:

$$\delta_c : \quad a_c(\delta_c(x,y),x,y) = 1. \quad (19)$$

In the ZEL case this is:

$$\delta_c(x,y) = 3 - (x + 2y). \quad (20)$$

This $\delta_c(x,y)$ function corresponds to the PS parameter $\delta_c$: much more dynamical information is present now. As expected, the shear lowers the value of $\delta_c$, i.e. it helps the collapse: with an adequate shear even underdense perturbations can collapse. Moreover, rare density peaks, which are great overdensities with respect to the density variance, are in general characterized by a low shear, so that their collapse is nearly spherical; this is in agreement with Bernardeau (1994).

In the case of spherical collapse, ZEL predicts $\delta_c = 3$, at variance with the exact value 1.69. This is not unexpected, as ZEL is known to underestimate the true collapse, especially in the spherical case. For the moment, let's assume that an "exact" $\delta_c$ function is given, in the sense that it correctly gives, as a function of the local variables $x$ and $y$, the initial density contrast necessary to make a perturbation collapse now. It is possible to write it as:

$$\delta_c(x,y) = \delta_0 - f(x,y) \quad (21)$$

where $\delta_0$ is the 1.69 spherical value, and $f(x,y)$ is a positive definite function, as the shear accelerates the collapse, with $f(0,0) = 0$.

As in the PS approach, it is assumed that a collapsing mass element, whose associated mass is a function of the mass variance $\sigma$ through the smoothing scale $R$, $M = M(R(\sigma))$, becomes part of a structure whose mass is greater than $M$. The cumulative MF, i.e. the fraction of matter contained in structures with mass greater than $M$, is then simply the integral of the PDF, $\mathcal{P}(\delta,x,y)$, over all the initial conditions which predict the perturbation to have collapsed within the present time, i.e.:

$$F(>M) = \int_0^\infty dx \int_0^\infty dy \int_{\delta_c(x,y)}^\infty d\delta\, \mathcal{P}(\delta,x,y). \quad (22)$$

The mass function is then:

$$N(M)dM = -\bar{\varrho}\frac{\partial}{\partial M}F(>M)\frac{dM}{M} = -\bar{\varrho}\frac{\partial F}{\partial \sigma}\frac{d\sigma}{dM}\frac{dM}{M}. \quad (23)$$

In this equation, the term which contains the dynamical information is $\partial F/\partial \sigma$. The integral of the PDF over $\delta$, being $\mathcal{P}(\delta)$ Gaussian, gives simply a term proportional to $\mathrm{erfc}(\delta_c(x,y)/\sqrt{2}\sigma)$, where erfc is the usual complementary error function. Integrating Eq. (22) in the variables $X \equiv x/\sigma$ and $Y \equiv y/\sigma$, the only $\sigma$ dependence of the cumulative MF $F$ is in the erfc term. We have:

$$\frac{\partial}{\partial \sigma}\mathrm{erfc}\left(\frac{\delta_c(\sigma X, \sigma Y)}{\sqrt{2}\sigma}\right) = -\frac{1}{\sqrt{2\pi}\sigma^2}\exp\left(-\frac{\delta_c^2}{2\sigma^2}\right)\left[\delta_c - \sigma X\frac{\partial}{\partial(\sigma X)}\delta_c - \sigma Y\frac{\partial}{\partial(\sigma Y)}\delta_c\right] \quad (24)$$

In the PS case the term $\partial F/\partial\sigma$ is:

$$\left.\frac{\partial F}{\partial \sigma}\right|_{PS} = \frac{\delta_0}{\sqrt{2\pi\sigma^2}} \exp\left(-\frac{\delta_0^2}{2\sigma^2}\right). \tag{25}$$

With Eqs. (17, 21, 22, 24, 25), the dynamical term $\partial F/\partial\sigma$ can be written as:

$$\frac{\partial F}{\partial \sigma} = \left.\frac{\partial F}{\partial \sigma}\right|_{PS} \times \mathcal{I}(\sigma), \tag{26}$$

where $\mathcal{I}(\sigma)$ is a correction factor to the PS mass function which gives the contribution to a given mass scale of non-spherically collapsing clumps:

$$\mathcal{I}(\sigma) \equiv \int_0^\infty dX \int_0^\infty dY\, \mathcal{P}(X,Y) \exp\left(-\frac{1}{2\sigma^2}f^2(\sigma X, \sigma Y) + \frac{\delta_0}{\sigma^2}f(\sigma X, \sigma Y)\right) \left[1 - \frac{1}{\delta_0}g(\sigma X, \sigma Y)\right], \tag{27}$$

where $g(x,y) \equiv f - x\partial f/\partial x - y\partial f/\partial y$. This expression is very general: given any local estimate of the collapse time, a function $\delta_c$ can be constructed by means of Eq. (19); then the corresponding correction factor $\mathcal{I}(\sigma)$ to the PS mass function is given by Eq. (27).

A remark on the normalization: the fact that the PS normalization is wrong by a factor $2=1/0.5$ is simply due to the fact that only overdense regions, which account for 0.5 of the mass, are predicted by the spherical model to collapse at some time. In the case of ZEL approximation, all the mass elements with negative $\lambda_3$-values are going to collapse at some time, and this happens for 92% of the mass (Doroshkevich 1970). It could then be attempted to get the right normalization by multiplying the ZEL-based MF by a factor $1/0.92 \simeq 1.09$. However, if excursion set theory has justified the factor $1/0.5$ for the PS mass function, it is not guaranteed that it would justify an analogous one $1/0.92$ for the ZEL case. As a matter of fact, we are interested in the large-mass part of the MF, where a PS-like approach is more believable, and whose structures collapse in a quasi-spherical way, the case in which the fudge factor 2 is justified. Then the MF has been normalized so as to recover the correctly normalized PS one in the spherical case, which is $\delta_c = \delta_0$, $\mathcal{I}(\sigma) = 1$, i.e. the PS and corrected MFs have been multiplied both by the same fudge factor 2. This causes the corrected MF to have a wrong normalization by a factor 1.84, which is due presumably to an overcounting of objects which collapse in a definitely non-spherical way, i.e. small-mass objects.

## 3.2 Realistic Estimates of the Collapse Time

As stated before, the plain ZEL prediction for the $\delta_c$ function does not reduce to the correct spherical value in the zero-shear case. Fig. 1 shows the ZEL $\delta_c$ curve, Eq. (20) (dashed line), as a function of the variable $z = x + 2y$. A simple way to force $\delta_c$ to reach $\delta_0=1.69$ at $z=0$ is to join it with the $(0,\delta_0)$ point by means of a horizontal branch: $\delta_c = \min\{\delta_0, 3 - z\}$. Physically, the mass element is supposed to follow spherical collapse, unless plain ZEL prediction allows it to collapse faster. Other possible $\delta_c$ curves, which have the same normalization, can be obtained by shifting the plain ZEL prediction horizontally, always connecting it to $\delta_0$ with a horizontal line (dotted lines in Fig. 1). This can be expressed as a two-parameter *ansätz* for the function $\delta_c$:

$$\delta_c = \min\{\delta_0, \delta_1 - z\}; \tag{28}$$

it will be referred to as *ansätz1*. The parameter $\delta_1$ describes how ZEL underestimates ($\delta_1<3$) or overestimates ($\delta_1>3$) the collapse rate of a mass element. In the limiting case $\delta_1=\delta_0$, the plain ZEL curve is just shifted to cross the $\delta_c$ axis at $\delta_0$. The PS case is also recovered in the limit $\delta_1 \to \infty$. Inserting Eq. (28) into Eq. (27), $\mathcal{I}(\sigma)$ can be written as:

$$\mathcal{I}(\sigma) = \iint_{Z \leq \Delta/\sigma} dX\, dY\, \mathcal{P}(X,Y) + \frac{\delta_1}{\delta_0} \exp\left(-\frac{\delta_1^2 - \delta_0^2}{2\sigma^2}\right) \iint_{Z \geq \Delta/\sigma} dX\, dY\, \mathcal{P}(X,Y) \exp\left(\frac{\delta_1}{\sigma}Z - \frac{1}{2}Z^2\right). \tag{29}$$

where $Z = z/\sigma$, and $\Delta = \delta_1 - \delta_0$. Fig. 2 shows some $\mathcal{I}(\sigma)$ curves for $\delta_1$-values 2, 2.5, 3 and 3.5 (dashed lines).

It is easy to obtain the limiting behaviours of Eq. (29). In the limit of large mass variances, which corresponds to low masses, $\Delta/\sigma$ is small, and then the first integral gives negligible contribution, while the second one turns out to be a constant, whose value is $\mathcal{I}(\sigma) \to 0.33\,\delta_1/\delta_0$. In the limit of small variances, the second integral can be calculated in the variables $X$ and $Z$:

$$\mathcal{I}(\sigma) \to \frac{20\sqrt{5}}{81\sigma^2} \frac{\delta_1^3}{\delta_0} \exp\left(\frac{9\delta_0^2 - 5\delta_1^2}{18\sigma^2}\right) \tag{30}$$

The sign of the constant $(9\delta_0^2 - 5\delta_1^2)$ determines whether $\mathcal{I}(\sigma)$ continues growing at small $\sigma$ or has a maximum and then falls down. The former behaviour is present when $\delta_1 < \sqrt{9/5}\delta_0 \simeq 2.27$; in this case the exponential cutoff of the $\partial F/\partial \sigma$ term, Eq. (26), becomes $\exp[-5\delta_1^2/18\sigma^2]$, instead of the PS $\exp[-\delta_0^2/2\sigma^2]$ one. In the $\delta_1 > \sqrt{9/5}\delta_0 \simeq 2.27$ case, the second integral of Eq (29) becomes negligible at large $\sigma$, while the first one approaches unity (for large $\Delta/\sigma$ that is just the integral of a normalized probability), so that $\mathcal{I}(\sigma) \simeq 1$. It is remarkable that, while the collapse of rare events is always nearly spherical, it does not guarantee the PS mass function to be recovered at large masses.

In Fig. 2 the limiting behaviour, Eq. (30), is reported (dotted lines); it describes very well the relevant small-$\sigma$ features of the $\mathcal{I}(\sigma)$ curves. In this way it is possible to estimate the position of the maximum of $\mathcal{I}(\sigma)$ in the $\delta_1 > \sqrt{9/5}\delta_0$ case, which is

$$\sigma_{MAX} \simeq \sqrt{(5\delta_1^2 - 9\delta_0^2)/18}, \tag{31}$$

and its height,

$$\mathcal{I}_{MAX} \simeq 3.66 \frac{\delta_1^3}{\delta_0(5\delta_1^2 - 9\delta_0^2)}. \tag{32}$$

In this *ansätz1*, being the $\delta_c$ curve flat around $z = 0$, perturbations with a small initial shear collapse exactly as spherical ones. As a consequence, the MF reduces to the PS one ($\mathcal{I}(\sigma) \to 1$), unless the contribution from non-spherically collapsing perturbations, which is given by the second integral, overcomes the one from the quasi-spherical clumps. This is not realistic, as it would mean that the largest collapsed structures are not roughly spherical, like rich clusters, but rather elongated ones, like collapsed filaments.

It is interesting to analyze what happens if the $\delta_c$ curve is not exactly flat around $z = 0$, a realistic possibility. Let us consider a second *ansätz* for the function $\delta_c$:

$$\delta_c(z) = \delta_0 - \delta_2' z \tag{33}$$

where $\delta_2'$ is the slope of the function $\delta_c$ at small $z$ (Fig. 1, dot-dashed lines, $\delta_2'=0.1$, 0.2, 0.3); PS is recovered for $\delta_2'=0$. This *ansätz* will be referred to as *ansätz2*. Fig. 3 shows the resulting $\mathcal{I}(\sigma)$ curves for $\delta_2'=0.1$, 0.2 and 0.3 (dashed lines). Putting Eq. (33) into Eq. (27), and integrating in the variables X and Z, in the small $\sigma$ limit one obtains:

$$\mathcal{I}(\sigma) \to \frac{60\sqrt{5}}{\sigma^2} \frac{\delta_2'^2 \delta_0^2}{(5 + 4\delta_2'^2)^{5/2}} \exp\left(\frac{\delta_0^2}{2\sigma^2}\xi\right), \tag{34}$$

where $\xi \equiv 4\delta_2'^2/(5 + 4\delta_2'^2)$. In this way the PS exponential cutoff becomes $\exp[-\delta_0^2(1 - \xi)/2\sigma^2]$, so that the effective value of $\delta_0$ is lowered by a factor $\sqrt{1 - \xi}$. It is interesting to note that, even for vanishing but non-null $\delta_2'$-values, the $\mathcal{I}(\sigma)$ function diverges at small $\sigma$ instead of approaching 1. This could seem in contradiction with the fact that the collapse of rare event is nearly spherical. However, the physical quantity is not $\mathcal{I}(\sigma)$ but the MF, and an exponential divergence of $\mathcal{I}(\sigma)$ at small $\sigma$-values corresponds to an effective change of $\delta_0$, which is vanishingly small when $\delta_2'$ approaches zero. Again, the fact that rare events collapse in a quasi-spherical way does not assure the PS to be recovered at large masses, but this time the added objects are quasi-spherical.

To give more realistic and dynamically motivated expressions for the function $\delta_c(x, y)$, two different models are proposed. First, in the evolution equation for the density, Eq. (5), the non-local shear-term can be expressed locally by means of ZEL (Eq. 11). Then Eq. (5) can be solved numerically to give an estimate of the collapse time; this procedure will be referred to as ZEL model. It is easy to see that, as long as ZEL predicts a slower collapse than the spherical case, ZEL shear has not a strong effect and then the density diverges nearly at the time predicted by the spherical model, while in the opposite case the divergence of ZEL shear drives the collapse, that will then happen nearly at the ZEL-predicted time. In practice, ZEL model is a dynamically motivated way to continuously join the spherical and ZEL regimes together. Inverting Eq. (19) it is possible to find the corresponding $\delta_c$ function, which is shown in Fig. 1 (heavy dashed line); note however that this time $\delta_c$ is not simply a function of $z = x + 2y$, except at large shears, but as a matter of fact the difference from the plain $z$-dependence is small; the curve in Fig. 1 is representative of the mean behaviour. As expected, ZEL model is very similar to the *ansätz1* with $\delta_1=3$, with a non-vanishing slope at small $x$ and $y$. The ZEL model $\mathcal{I}(\sigma)$ curve is shown in Fig. 2 (heavy line); again it is very similar to the *ansätz1* $\delta_1=3$ curve, apart from the weak exponential rise at small $\sigma$, which takes place when the PS mass function has already been cut off. As a consequence, all the features of *ansätz1* $\delta_1=3$ described before (Eqs. 31 and 32) can be applied to ZEL model to quantify its relevant features.

The second model is the well-known homogeneous ellipsoid model (e.g., Peebles 1980); it will be called ELL model. Care has to be used in introducing it because, while a spherical perturbation can be consistently put in an otherwise homogeneous Universe, this is not possible for a triaxial overdensity. Again, what has to be done is to "extract" an ellipsoid from a whole perturbed Universe. Notably, the dynamical equations for the evolution of an ellipsoid are formally the same as the ones for the evolution of a general Lagrangian mass element (Peebles 1980); what is totally different is the role of the potential, which is what gives informations about the rest of the world. Following Bond & Myers (1993a), the gravitational potential can be expanded in a Taylor series:

$$\phi(\mathbf{x}) = \phi_0 + \frac{\partial \phi}{\partial x_i} x_i + \frac{1}{2} \frac{\partial^2 \phi}{\partial x_i \partial x_j} x_i x_j + \ldots \qquad (35)$$

The first term is a constant and has no importance; the second term induces just a bulk motion on the mass element. The third term, which is a quadratic form and contains the Laplacian of the potential (its trace) and the tidal field (its traceless part), is the first and only relevant term which influences the internal properties of a mass element (no further derivatives of the potential enter the Lagrangian evolution equations of a mass element, Eqs. 1, 2 and 3). This is splitted in two parts:

$$\phi = \phi_{EXT} + \phi_{INT}. \qquad (36)$$

The first term, traceless, describes the external tides and is dominant in the early evolution; the other term, which dominates in the non-linear phase, describes internal dynamics, and can be written as the potential of a homogeneous ellipsoid, which is possible as it is a quadratic form. Initial conditions are set up so as to be consistent with ZEL at early times; again they are given by the three eigenvalues $\lambda_i$.

The equations of motion of the ellipsoid, which are reported in Appendix A, have been numerically integrated to determine the collapse time. Consistently with what has been done before, the collapse time has been defined as the instant at which the density becomes very large, which happens when the ellipsoid collapses along its shortest axis; this point will be discussed at the end of §4. The resulting $\delta_c$ function is shown in Fig. 1 (heavy continuous line); again, the curve in Fig. 1 is just representative, as $\delta_c$ is not simply a function of $x + 2y$, but the difference from this behaviour is not strong. This time the slope of $\delta_c$ at small $x$ and $y$ is larger than the previous cases; moreover, Ell model predicts slower large-shear collapses than the pure ZEL prediction, which is probably due to the strong non-radial motions which are important in these cases. Fig. 3 shows the resulting $\mathcal{I}(\sigma)$ function (heavy line), in which the exponential rise at small variances is apparent; *ansätz2* curves with a $\delta'_2$ parameter around 0.2 give a qualitatively good fit. As the asymptotical behaviour of ELL model at large $x$ and $y$ has turned out difficult to determine, and given that only the small-shear part is relevant for our purposes, plain ZEL prediction has been forced after $x + 2y = 4$; this causes the small bump in the ELL $\mathcal{I}(\sigma)$ curve at $\sigma \sim 4$ and determines the asymptotical large-variance behaviour, which is the same as the ZEL model one. The pure exponential behaviour of *ansätz2* $\mathcal{I}(\sigma)$ takes place at very small $\sigma$-values. It is then useful to compare the ELL $\mathcal{I}(\sigma)$ function directly to the one necessary to make the MF be equal to the PS one with an effective $\delta_0$ value, say $\delta_E$, instead of the 1.69 one:

$$\mathcal{I}(\sigma) = \frac{\delta_E}{\delta_0} \exp\left(\frac{\delta_0^2 - \delta_E^2}{2\sigma^2}\right). \tag{37}$$

Fig. 3 shows the $\mathcal{I}(\sigma)$ curves corresponding to $\delta_E$=1.6, 1.5 and 1.4, values which have been found to fit several N-body results. The fit with the ELL curve is only rough, but considering that the small-$\sigma$ difference can be lost in N-body simulations because of limited statistics, and that the $\sigma \sim 1$ difference, which is slightly more than a factor 2, can be lowered by a correct normalization procedure (non-spherical objects are overcounted) and by fragmentation of clumps, it is reasonable to think that such an ELL MF could be roughly fit by a $\delta_E < \delta_0$ PS curve.

## 4  Results and Discussion

The last quantity needed to calculate the MF (Eq. 23) is the cosmological term $d\sigma/dM$; it introduces the dependence of the MF on the power spectrum of perturbations. It is useful to consider first scale-invariant power-law power spectra: $P(k) \propto k^n$. In this case:

$$\sigma(M) = (M/M_0)^{-(3+n)/6} \tag{38}$$

where $M_0$ is the mass scale at which $\sigma = 1$. Then our new MF takes the form:

$$N(M) = \frac{\bar{\varrho}}{\sqrt{\pi}} \left(1 + \frac{n}{3}\right) \left(\frac{M}{M_*}\right)^{\frac{3+n}{6} - 2} \exp\left(-\left(\frac{M}{M_*}\right)^{\frac{3+n}{3}}\right) \mathcal{I}\left(\frac{\delta_0}{\sqrt{2}} \left(\frac{M}{M_*}\right)^{\frac{3+n}{6}}\right) \frac{1}{M_*^2} \tag{39}$$

where $M_* = M_0(2/\delta_0^2)^{3/(3+n)}$. As expected, the corrected MF is scale invariant as long as the spectrum is scale invariant. Spectra have been normalized by the relation $\sigma(8h^{-1}\text{Mpc})=1$ (on a top-hat sphere) with $h$=0.5, and gaussian smoothing is assumed. Fig. 4 shows the results for the PS (dashed), $\delta_E$=1.5 PS (dotted), ZEL (dot-dashed) and ELL (continuous) MFs, for spectral indexes $n$ from $-2$ to 1. The enhancement of the large-mass part is apparent in all the cases; in particular, the ZEL MF predicts more structures than the PS one around $M_*$ but reduces to it at large masses, while the ELL MF is has its large-mass tail shifted toward large masses, and is well-fitted in that part by the $\delta_E$=1.5 curve.

Because of Eq. (38), mass scales are more diluted for small spectral indexes, so that both the extent of the bump in the ZEL case and the shift of the large-mass part in the ELL case are larger for $n = -2$. With the aid of the two *ansätz* presented in the previous section, it is possible to quantify in a more precise way the two features. Using *ansätz*1 with $\delta_1 = 3$ for the ZEL $\mathcal{I}(\sigma)$ function, the maximum of the bump is at $M_{MAX} = (\delta_0^2/5)^{3/(3+n)} M_* \simeq 0.57^{3/(3+n)} M_*$ (Eq. 31) and its maximum height is a factor $\simeq 3.03$ (Eq. 32). For the ELL mass function, it suffices to say that the shift of $M_*$ is $(\delta_0/\delta_E)^{6/(3+n)}$ (see Eq. 37), which is larger for smaller spectral indexes.

All these features are recovered in more realistic power spectra. As an example, the standard cold dark matter (CDM) model has been used. The transfer function has been taken from Bardeen et al. (1986); $\Omega = 1$, $h = 0.5$ and $b = 1$ have been assumed; gaussian smoothing has been used; the normalization has been fixed by $\sigma_{TH}(8h^{-1}Mpc) = 1$, which is roughly consistent with COBE, given $b = 1$. Fig. 5 shows the four MFs (PS, $\delta_E$=1.5 PS, ZEL and ELL) at different times; as expected, at earlier times, when the logarithmic slope of the spectrum at the collapsing scale is lower, the ZEL bump is slightly more diluted and the ELL shift of $M_*$ is slightly more pronounced. In any case, a rather firm conclusion can be drawn: because of the fact that non-spherical collapses are faster than spherical ones, more large-mass objects are present with respect to the PS mass function. This is consistent with the findings of Lucchin & Matarrese (1988), who have found an increase of $M_*$ due to the non-Gaussianity of the evolved matter field, and with Cavaliere & Menci (1994), who have found a similar effect by taking into account the coagulation of clumps.

The validity of the PS-like approach we have used is limited to the large-mass part of the MF ($M > M_*$) for a number of reasons. Single-stream dynamics has been assumed, which is valid if a mass element has a small probability of residing in multi-stream regions; Kofman et al. (1994) have shown that multi-stream regions are rare for mass variances smaller than $\sigma \sim 1$, which means that a PS-like approach can be meaningful only for masses larger than $M_*$. Furthermore, coagulation and fragmentation of clumps have been neglected. It is reasonable to assume that coagulation is negligible in single-stream regime (clumps have to come in touch to coagulate), while fragmentation of clumps is expected to be negligible when a mass element collapses in a quasi-spherical way, so that tides can not be so efficient to disrupt it. Another assumption is that the collapse time could be predicted from local initial conditions. Notably, a mass element spends most of its time before collapse in the linear and quasi-linear regime, where single-stream regions dominate and ZEL description is accurate, while the following highly non-linear phases are much faster. As a consequence, the collapse time predictions that have been presented here are likely to be realistic, while subsequent fragmentation and coagulation histories can give severe problems at small masses. A last assumption, already discussed before, is that PS normalization can be applied. This is surely wrong at small masses by a factor 1.84, but is probably accurate in the large-mass part, where the collapses are quasi-spherical. The application of excursion set theory can solve this last problem, but this time the diffusion of trajectories in the $\lambda_1$, $\lambda_2$ and $\lambda_3$ (or $\delta$, $x$ and $y$) space has to be considered, and the whole $\delta_c(x,y)$ function has to be used as an absorbing barrier.

There is a number of points which deserve further discussion. On the actuality of Bertschinger & Jain's collapse theorem, which states that the spherical perturbation is the slowest to collapse, there is not full agreement in literature. In practice, that theorem is valid for a fluid element, which has no structure by definition, and is then appropriate for our approach, in which the power spectrum has to be truncated to avoid strong non-linearities. Different reasonings, which can end up in different conclusions, apply to an extended structure. For instance, Peebles (1990) and Bond & Myers (1993a) claim the shear to slow-down the collapse. The former's result will be discussed below; the latters, applying the ELL model to the collapse of density peaks to determine the instant at which full virialization is reached, have chosen to stop the evolution of every collapsing axis of the ellipsoid at a certain point, mimicking in this way a process of partial virialization, and to wait for the collapse of all the axes to define the collapse time. As a consequence, a non-spherical collapse is slower than the corresponding zero-shear spherical one. Actually, after first-axis collapse, because of partial virialization, matter can not be assumed any more cold and in single-stream regime, and then ELL model ought not to be extrapolated beyond that moment; nonetheless, predictions based on that model have been found in satisfactory agreement with N-body simulations (Bond & Myers 1993b).

In the present work, first-axis collapse has been chosen here as a suitable definition of collapse time. This different choice is mainly motivated by the fact that large overdensities and not fully virialized structures are searched. In practice, a mass element, as the ones that have been considered here, has no internal properties other than density, expansion and shear, so that the only meaningful definition of collapse is when those quantities reach large values; this happens at the moment corresponding to the first-axis collapse. It has been often said that such a collapse is only kinematical, in the sense that matter collides because of its initial velocity, while the internal potential does not diverge (Shandarin & Zel'dovich 1989). On the other hand, the first-axis collapse is accompanied by shock waves in the baryonic component (Zel'dovich 1970), and after that moment matter can not be described any more as a cold fluid in single-stream regime, so that this is the true moment of transition to highly non-linear dynamics. Another objection which could be done is that in this way pancake-like objects are counted. It is to be stressed, however, that pancakes are only a transient feature of a self-gravitating system (Shandarin & Zel'dovich 1989), which disappears after the collapse. So, a collapsing mass element will only show a filamentary phase in the first instants of collapse, as is seen in N-body simulation of clusters (e.g., White 1994), but will then become a clumpy object; the subsequent possible fragmentation of the clump is the really worrying feature.

A last assumption which has been made is that small-scale structure can be safely neglected by truncating the power spectrum at some wavelength. Taking small-scale structure into account, arguments based on least-action calculations (Peebles 1990) and dynamical friction (Antonuccio & Colafrancesco 1993) lead to the conclusion that a collapsing structure is severely slowed down just after it has detached from Hubble expansion. Peebles (1990) has also shown that the presence of external tides induces strong non-radial flows in an otherwise spherical overdensity, so that the collapse is slowed down and then non-spherical collapses are slower than spherical ones, a conclusion which is again opposite to Bertschinger & Jain's collapse theorem. By the way, it is to be noted that Peebles' and Bertschinger & Jain's statements have been obtained in quite different frameworks, so that a comparison of the two is not straightforward. Nonetheless, it is to be retained that, as more small-scale power is present, the collapse of a perturbation is slowed down in a way that could inhibit the positive effect of the shear.

# 5 Summary and Conclusions

In this paper the original Press & Schechter (1974) approach to the mass function has been extended to non-spherical dynamics. Dynamical approximations and models of non-spherical collapse have been used to determine the instant of collapse of a Lagrangian mass element, starting from the local eigenvalues of the deformation tensor, which are the suitable initial conditions for local dynamical predictions. The Zel'dovich (ZEL) approximation, which is a true self-consistent first-order Lagrangian and local approximation, has been used as a guide for the determination of the collapse time. As ZEL does not reduce to the exact spherical model in the zero-shear case, two *ansätz* for the true collapse time have been analysed in order to understand some relevant features. For a realistic estimate of the collapse time, two models have been used, one based on the numerical solution of the exact Lagrangian equation for the density contrast, but with the shear given by ZEL (ZEL model), the other based on the numerical solution of the homogeneous ellipsoid model up to the first-axis collapse (ELL model). Two fundamental merits of these ZEL inspired models, with respect to the spherical collapse, are the fact that collapsing clumps are consistently extracted from a perturbed field, instead of being put in an otherwise homogeneous Universe, and that the shear of the perturbation, which is the kinematical quantity that is missed by the spherical model, is taken into account.

As a result, the collapse time of general non-spherical perturbations is shortened, in agreement with Bertschinger & Jain's collapse theorem, and then more large-mass objects are formed; ZEL model predicts more objects to be present around the knee of the MF, the ELL model, which is probably more realistic, is well fitted in the large-mass end by a PS function with a lower value of the effective $\delta_c$ parameter, in the range $1.4 \div 1.6$. This last prediction is in agreement with many N-body simulations: e.g. Efstathiou & Rees (1988) have found $\delta_c = 1.33$, Carlberg & Couchman (1989) 1.44, Bond & Myers (1993b) 1.58, Klypin et al. (1994) 1.33, while Efstathiou et al. (1988), Bond et al. (1991) and Lacey & Cole (1994) have used the 1.69 spherical value. However, in all those cases the PS mass function has been treated as a phenomenological function, and no dynamical interpretation has never been given for $\delta_c$-values smaller than 1.69.

When small-scale structure is present, dynamical friction and previrialization events can slow down the collapse rate of a perturbation, thus increasing the effective value of the $\delta_c$ parameter. As a matter of fact, Klypin et al. (1994) have found their N-body MFs to be fitted at different times by PS functions with different $\delta_c$-values, larger at later times. However, this result could be caused, at least in part, by their clump-finding algorithm, which can underestimate the number of large-mass objects. As mentioned in the introduction, also Jain & Bertschinger (1993) have found their N-body MFs to give more large-mass objects than the pure PS prediction at large redshifts, consistently with their second-order calculation of the power spectrum, while at later times the PS curve has been found to give a good fit and even overestimate the N-body curve. In their case, the earlier-time N-body MFs are not well fitted by PS with a smaller $\delta_c$. It could be speculated then that the restoration of PS shape at later time is a consequence of the fact that previrialization tends to contrast the positive effect of the shear, so that at a certain moment spherical and non-spherical perturbations can collapse at the same time.

To conclude, the PS mass function with $\delta_c$-values in the range $1.4 \div 1.6$ is confirmed to give a good first approximation of the MF of large cosmological structures; besides, the use of non-spherical collapse models gives a dynamical explanation of the lower $\delta_c$ parameters used to fit some N-body simulations, though the true shape of the mass function can be somehow different from the PS one. The possible decrease with time of the number of large-mass objects, or equivalently the increase of the effective $\delta_c$ at later times, can have interesting consequences on the abundance of objects at high redshifts, as quasars (Efstathiou & Rees 1988) or rich clusters of galaxies (Evrard 1989; Peebles, Daly & Juszkiewicz 1989). In particular, it could explain, at least in part, why rich clusters can be found at very high redshifts, which seems incompatible with their abundance in the near Universe.

The author thanks Sabino Matarrese for his help and encouragement, and Stefano Borgani, Marco Bruni, Paolo Catelan, Giuliano Giuricin, Fabio Mardirossian, Nicola Menci and Marino Mezzetti for enlightening discussions.

## Appendix A: Equations for the Homogeneous Ellipsoid Model

In this appendix the evolution equations for an homogeneous ellipsoid, ELL model, are given (for a derivation see Bond & Myers 1993a). Let $r_i(t) = a_i(t) q_i$ the physical $i$-th coordinate of the outer surface of the ellipsoid, in the principal-axes system; $q_i$ is the initial — Lagrangian — position and $a_i$ is the expansion factor for the $i$-th axis. The universal scale factor is called $\bar{a}$. Then the evolution equations for the $a_i$ factors are:

$$\frac{d^2 a_i}{dt^2} = -4\pi G \bar{\varrho} a_i \left[ \frac{1}{3} + \frac{\delta}{3} + \frac{b'_i}{2}\delta + \lambda'_{vi} \right] \tag{40}$$

where

$$\delta = \frac{\bar{a}^3}{a_1 a_2 a_3} - 1 \tag{41}$$

$$b'_i = \frac{2}{3}[a_i a_j a_k R_D(a_i^2, a_j^2, a_k^2) - 1] \quad i \neq j \neq k \tag{42}$$

$$\lambda'_{vi} = -\frac{\bar{a}}{\bar{a}_0}\left(\frac{\delta}{3} + \bar{a}_0 \lambda_i\right) \tag{43}$$

Here the $\lambda_i$ parameters are the eigenvalues of the initial deformation tensor, $\partial u_i(a_0)/\partial q_i = \text{diag}(\lambda_1, \lambda_2, \lambda_3)$, and $R_D$ is the Carlson's elliptic integral:

$$R_D(x, y, z) = \frac{3}{2} \int_0^\infty \frac{d\tau}{(\tau + x)^{1/2}(\tau + y)^{1/2}(\tau + z)^{3/2}};  \qquad (44)$$

this integral has been calculated by means of the routine given by Press & Teukolsky (1990).

The initial conditions have been set up in order to be consistent with ZEL at early times:

$$a_i(t_0) = \bar{a}_0(1 + \bar{a}_0 \lambda_i) \qquad (45)$$

$$\frac{da_i}{dt}(t_0) = \frac{1}{\bar{a}_0}\left(\frac{d\bar{a}}{dt}\right)_0 \left(a_i(t_0) + \bar{a}_0^2 \lambda_i\right) \qquad (46)$$

Eq. (40) reduces to the usual Friedmann equation if all $\lambda_i = 0$ (which implies of course $\delta_0 = 0$). The second term between squared parentheses in the rhs of Eq. (40) gives the spherical contribution from the overdensity, the third and fourth ones give the contribution from internal and external tides. This last term is evolved according to linear theory (Eq. 43), which corresponds to holding the peculiar potential constant; this is a good approximation as long as the perturbation is in the quasi-linear regime (e.g., Bagla & Padmanabhan 1994), while in the highly non-linear regime the second and third term, the internal non-linear ones, dominate. Bond and Myers (1993a) have verified that changing this assumption with other reasonable ones does not affect the result.

# Figure captions

**Figure 1:** $\delta_c$ curves as a function of $z = x + 2y$: plain ZEL prediction (dashed), *ansätz1* curves (dotted), *ansätz2* curves (dot-dashed), ZEL model (heavy dashed) and ELL model (heavy continuous).

**Figure 2:** $\mathcal{I}(\sigma)$ curves of *ansätz1*, with $\delta_1$=2, 2.5, 3 and 3.5 (dashed lines); dotted lines represent the corresponding limiting behaviour, Eq. (30). Heavy continuous line: ZEL model.

**Figure 3:** $\mathcal{I}(\sigma)$ curves of *ansätz2*, with $\delta_2'$=0.1, 0.2, and 0.3 (dashed lines). Continuous lines correspond to PS curves (Eq. 37) with $\delta_E$=1.6 (left), 1.5 (center) and 1.4 (right). Heavy continuous line: ELL model.

**Figure 4:** The mass function for power-law spectra ($n = -2$ and 0): PS (dashed), PS with $\delta_E$=1.5 (dotted), ZEL model (dot-dashed) and ELL model (continuous).

**Figure 5:** The mass function for a standard CDM spectrum, at different times: PS (dashed), PS with $\delta_E$=1.5 (dotted), ZEL model (dot-dashed) and ELL model (continuous).

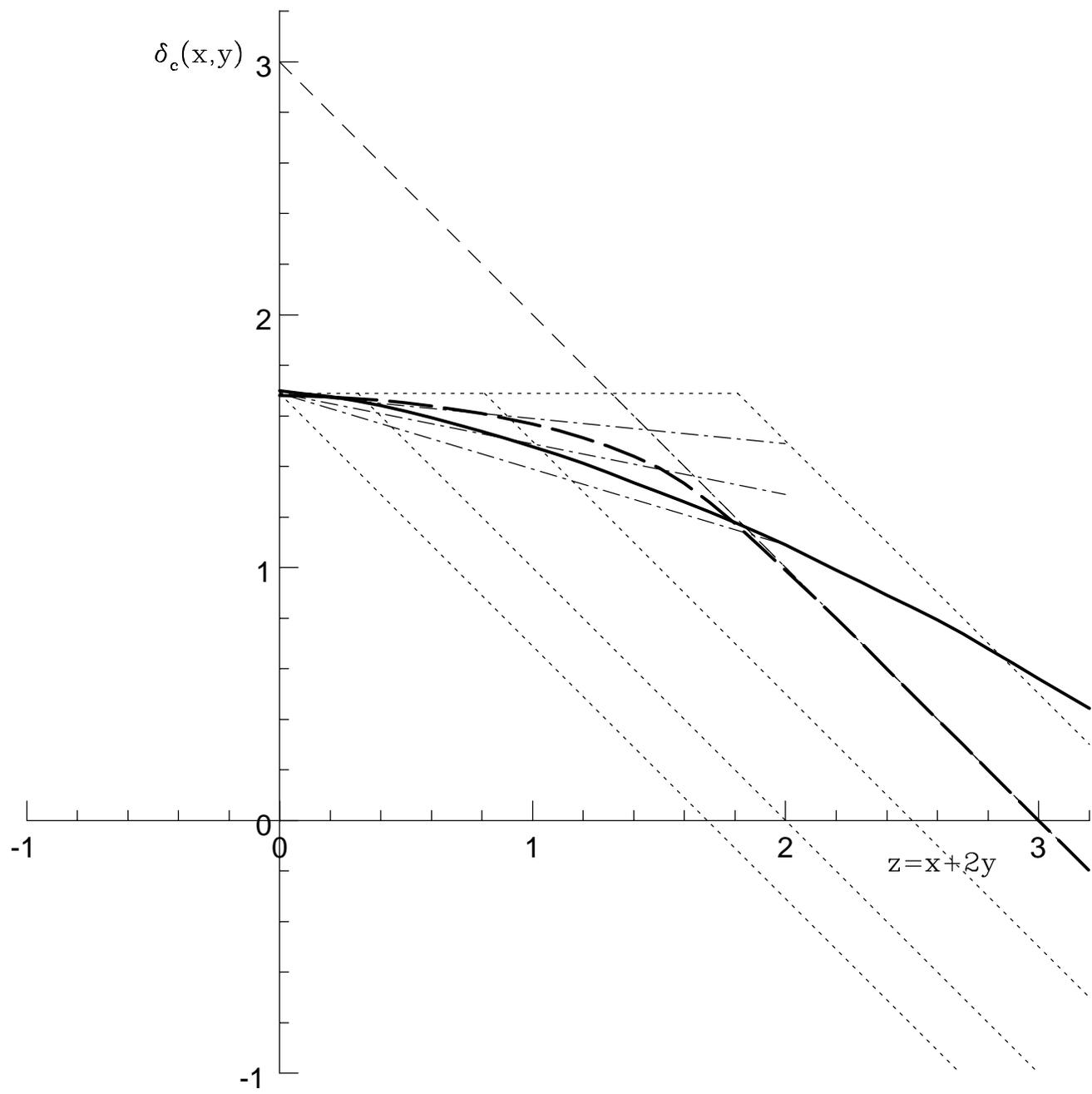

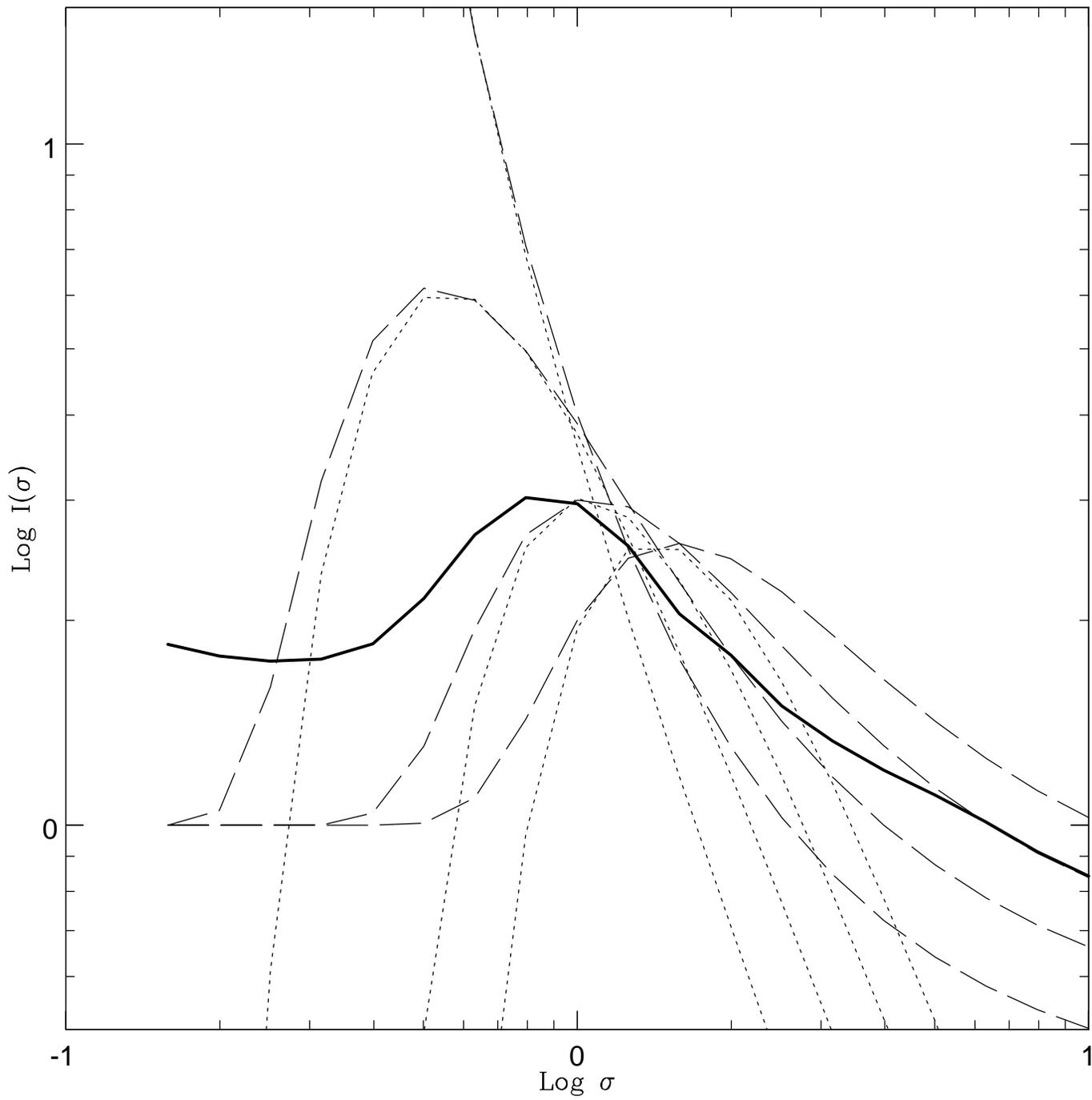

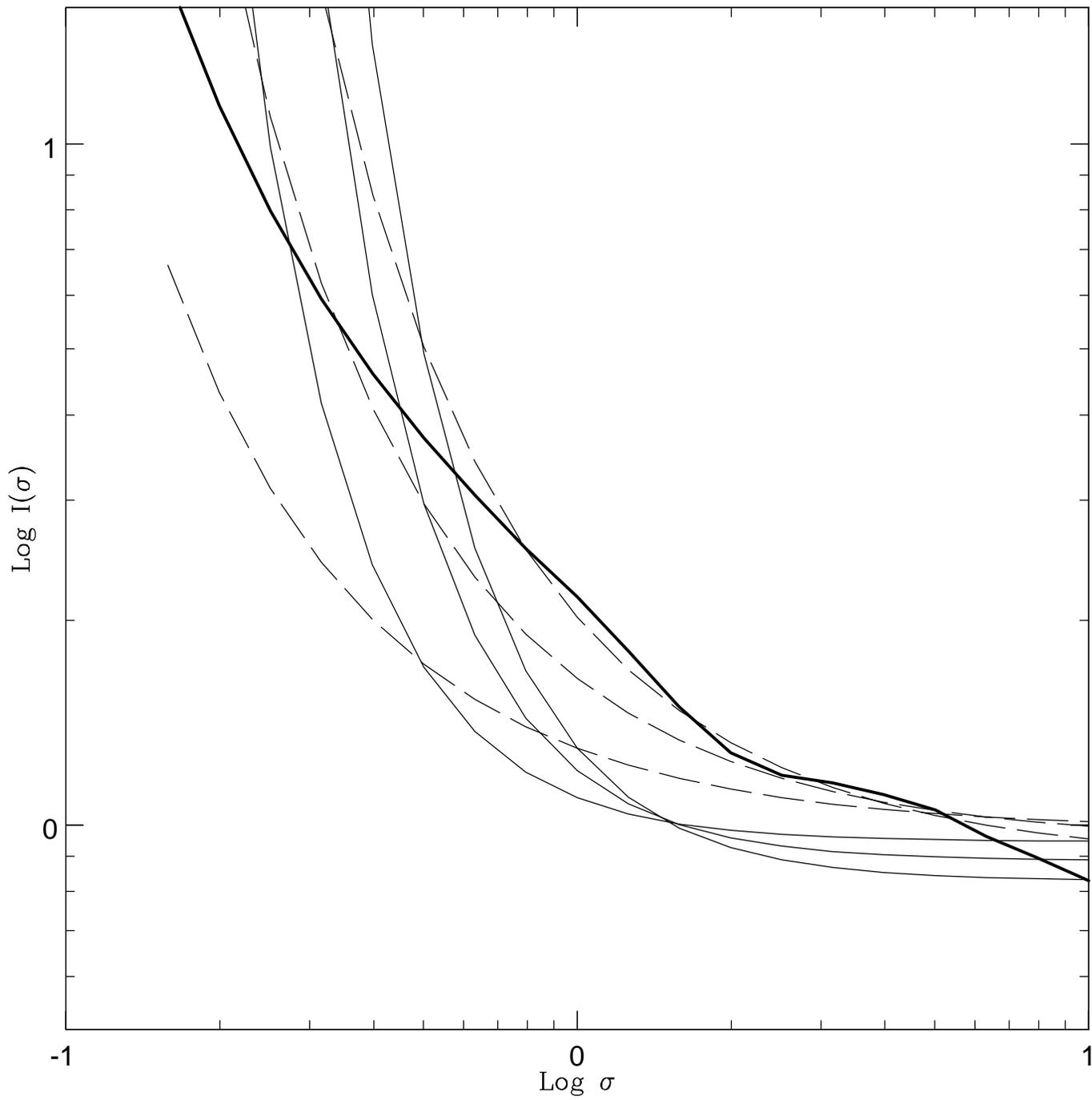

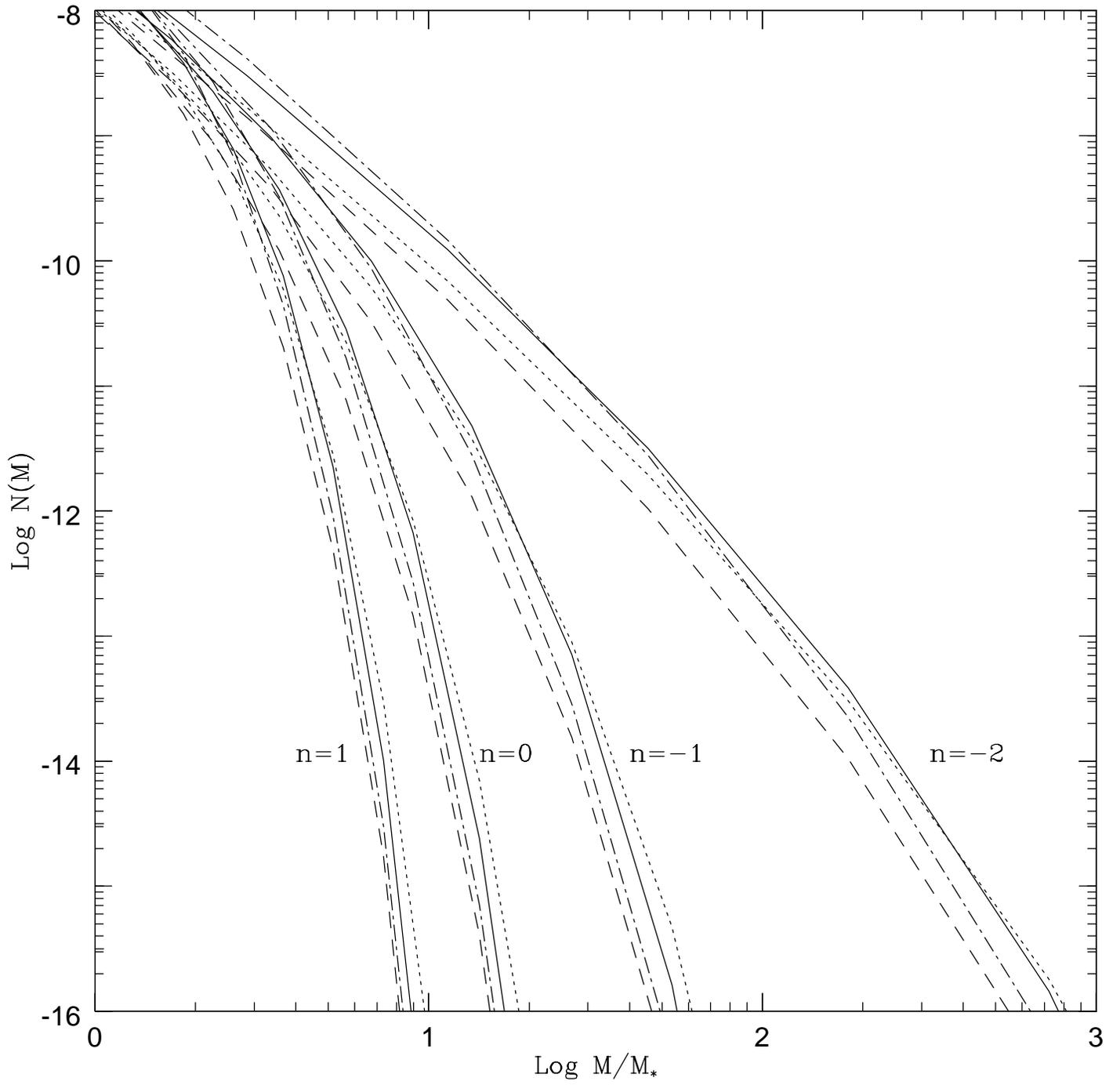

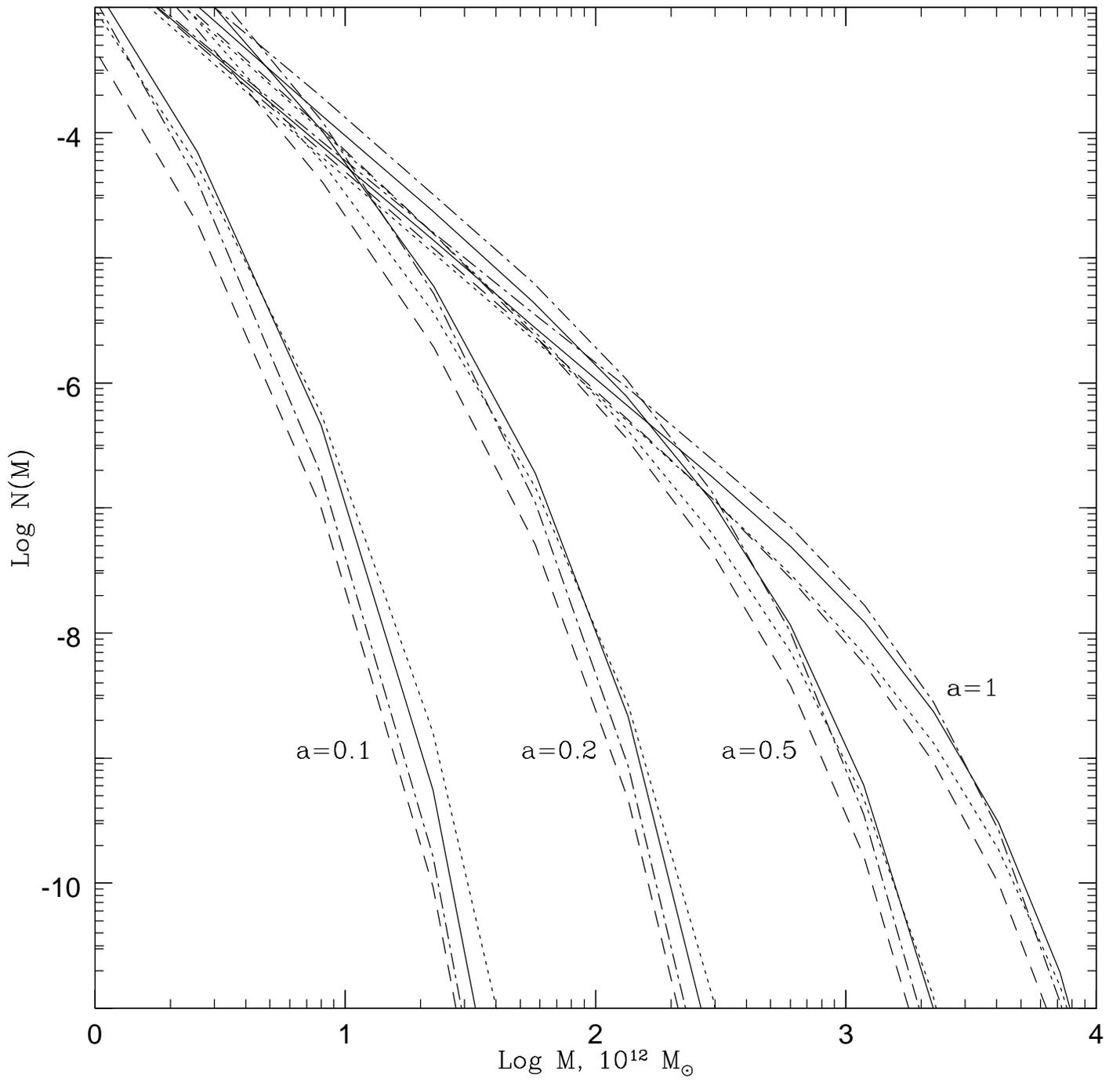